\documentclass{jpconf}

\usepackage{amsmath,amssymb}

\begin{document}

\title[Features of gravitational waves in higher dimensions]{Features of gravitational waves in higher dimensions}
\author{Otakar Sv\'{\i}tek}
\address{Institute of Theoretical Physics, Charles University in Prague, V Holesovickach 2, 180 00 Praha 8 - Holesovice, Czech Republic }
\ead{ota@matfyz.cz}
\begin{abstract}
There are several fundamental differences between four-dimensional and higher-dimensional gravitational waves, namely in the so called braneworld set-up. One of them is their asymptotic behavior within the Cauchy problem. This study is connected with the so called Hadamard problem, which aims at the question of Huygens principle validity. We investigate the effect of braneworld scenarios on the character of propagation of gravitational waves on FRW background.
\end{abstract}

\section{Introduction}

Gravitational waves in higher dimensions have several features that might distinguish them from the waves propagating on the four-dimensional manifold only. Therefore, many features were already studied, including quasi-normal modes, modification of late-time tails, and also the higher-dimensional exact radiative spacetimes.

Aside from these studies in general higher-dimensional spacetimes, special attention has been paid to specific braneworld models describing our universe as a hypersurface (brane) in a higher-dimensional spacetime (bulk). Apart from the quickly discovered short-distance modification of the Newtonian gravitational potential, a nontrivial correction to the quadrupole radiation formula has been derived \cite{kinoshita}. There are also the notorious KK modes that are effectively massive on the brane.

\section{Huygens principle and tails}
Huygens principle (and related Hadamard problem) deals with the dependence of field values on the Cauchy data. Namely, whether it suffices to know the data only in an arbitrarily small neighborhood of the intersection of the past null cone with the Cauchy surface. In numerous mathematical studies of the problem (namely works by Hadamard \cite{hadamard}, G\"{u}nther \cite{gunther} and McLenaghan \cite{mclenaghan}) it was shown that the result depends on both the differential operator properties and underlying spacetime. This makes it a possible candidate for studying deviations that might occur when we move to higher-dimensional spacetimes.

Related to the above described qualitative description there is a problem of presence and characteristic decay of late-time tails. It quantifies the fall-of of the initially sharp signal from distant source as seen by the observer at fixed position. In the context of physically interesting situations in four dimensions the study was started mainly by Price \cite{price}, in higher dimensions it was recently studied for example in \cite{rostworowski}.

The usual explanation of these late-time tails is the effect of backscaterring on the curvature (or the potential), but one should be careful to use this explanation for any conclusions.

\section{Brane model}
Our concern here is to analyze how the specific brane model affects the validity of Huygens principle of gravitational waves on FRW spacetime. In braneworld scenarios gravity feels the higher dimensional spacetime while other fields are confined only to four-dimensional brane. There are several different models and we will assume the Randall-Sundrum single brane set-up \cite{randall-sundrum}.

For the description of the model we use the Gaussian embedding technique \cite{shiromizu} with the following form of the metric
\begin{equation}
ds^2=dw^2+i_{\mu\nu}dx^{\mu}dx^{\nu}\ .
\end{equation}
We assume $Z_2$ symmetry with bulk cosmological constant. The five-dimensional Einstein equations have the form
\begin{equation}
G_{MN}=\kappa_{5}^{2}[-\Lambda_{5}g_{MN}+\tau_{\mu\nu}\delta_{M}^{\mu}\delta_{N}^{\nu} \delta(w)]\ .
\end{equation}
The matter content on the brane is given by the Israel junction conditions
\begin{equation}
K_{\mu\nu}^{+}=-K_{\mu\nu}^{-}=\kappa_{5}^{2}\left[\tau_{\mu\nu}-\frac{1}{3}i_{\mu\nu}(0)\tau^{\alpha}{}_{\alpha}\right]\ ,
\end{equation}
where the brane stress energy tensor is decomposed in the following way
\begin{equation}
\tau_{\mu\nu}=-\lambda i_{\mu\nu}(0)+S_{\mu\nu}\ .
\end{equation}
Four-dimensional Einstein equations then reduce to the form
\begin{equation}
{}^{4}G_{\mu\nu}=-\Lambda_{4}i_{\mu\nu}(0)+\kappa_{4}^{2}S_{\mu\nu}+\kappa_{5}^{4}\pi_{\mu\nu}+\bar{E}_{\mu\nu}(0)
\end{equation}
where $\Lambda_{4}=\frac{\kappa_{5}^{2}}{12}(\kappa_{5}^{2}\lambda^2+6\Lambda_{5})$, $\kappa_{4}^{2}=\frac{\kappa_{5}^{4}\lambda}{6}$, $\pi\sim S^2$, and $\bar{E}_{\mu\nu}(0)$ is a continuous part of certain Weyl tensor projection at the brane.

Since we would like to study cosmological model we assume the perfect fluid on the brane
\begin{equation}
S_{\mu\nu}=(p+\rho)u_{\mu}u_{\nu}+pi_{\mu\nu}\ .
\end{equation}
Then even $\pi_{\mu\nu}$ can be put to perfect fluid form with parameters quadratic in $p, \rho$.

\section{Modified Friedmann equation}
Restricting $i_{\mu\nu}(0)$ to the form of a standard FRW metric on the brane and assuming $\bar{E}_{00}(0)=\Lambda_{4}=0$ we arrive at the modified Friedmann equation corresponding to the previously derived results \cite{langlois}
\begin{equation}\label{mod-FRW}
\frac{3(\dot{a}^2+k)}{a^2}=\kappa_{4}^{2}\rho+\frac{\kappa_{5}^{4}\rho^2}{12}\ .
\end{equation}

For example the behavior of expansion function in the radiation era ($\rho=\rho_{0}a^{-4}$) with $k=0$ is the following
\begin{equation}
a(t)\sim t^{1/2}\left[1+\frac{\ell}{2t}\right]^{1/4}\ ,
\end{equation}
where the modification due to brane model is clearly visible as the second term in the bracket.

\section{Perturbations of the brane metric}
We consider FRW metric in the following form
\begin{equation}
ds^2=a^2(-d\eta^2+dr^2+\sigma^2 d\Omega^2)\ ,
\end{equation}
with conformal time $\eta$ and radial function $\sigma=r, \sin(r), \sinh(r)$ (depending on the spatial curvature $k=0,1,-1$).

Axial perturbations of gravitational field in the Regge-Wheeler gauge were derived in \cite{malec} and two polarizations might be parametrized as 
\begin{eqnarray}
h_{0\phi}&=&h_{0}(\eta,\rho)\sin(\theta)\ \partial_{\theta}Y_{lm}\nonumber\\
h_{1\phi}&=&h_{1}(\eta,\rho)\sin(\theta)\ \partial_{\theta}Y_{lm}
\end{eqnarray}
If compatibility condition
\[
(\sigma \sigma_{,r})_{,r}+2k\sigma^2-1=0
\]
is satisfied, then equations for both modes can be reduced to one equivalent scalar equation. Defining $F=\frac{h_{1}}{\sigma a}$ this equation is
\begin{equation}\label{eq-perturb}
\partial_{\eta}^2 F-\partial_{r}^2 F+\frac{l(l+1)}{\sigma^2} F-k F-\frac{\partial_{\eta}^{2}a}{a} F=0
\end{equation}

From the modified Friedmann equation \eqref{mod-FRW} with conformal time and assuming radiation era we get
\begin{equation}
\partial_{\eta}^{2}a=-ka+\frac{\kappa_{5}^{4}\rho^{2}a^{3}}{18}\ ,
\end{equation}
which after combining with \eqref{eq-perturb} leads to 
\begin{equation}\label{F-evolution}
\partial_{\eta}^2 F-\partial_{r}^2 F+\frac{l(l+1)}{\sigma^2} F-\frac{\kappa_{5}^{4}\rho^{2}a^{2}}{18} F=0\ .
\end{equation}
The behavior of the solutions to this equation will now be analyzed both in the case of standard and modified Friedmann evolution equations.

\section{Behavior analysis}
For standard cosmology the term $\sim \rho^2$ is missing in \eqref{F-evolution} and the solution is of the form
\begin{equation}
F_{l}\sim \sigma^{l}\underbrace{\partial_{r}\sigma^{-1}\cdots \partial_{r}\sigma^{-1}}_{\substack{l-1}} \partial_{r}\left(\frac{f+g}{\sigma}\right)\ ,
\end{equation}
with functions $f,g$ depending only on combinations $r-\eta$ and $r+\eta$ and so the gravitational radiation is bounded by null cones. Therefore, Huygens principle is valid in this case.

In our braneworld scenario the analysis is more difficult and the approach from \cite{malec} will be used. We start with the decomposition of solution $F=F_{0}+f$ , where $F_{0}$ satisfies the reduced equation (without the brane-induced modification) with initial data of compact support $\sigma(r)\leq\sigma(r_{0})$ and $f$ fulfills
\begin{equation}
\partial_{\eta}^2 f-\partial_{r}^2 f+\frac{l(l+1)}{\sigma^2} f=\frac{\kappa_{5}^{4}\rho^{2}a^{2}}{18}(f+F_{0})
\end{equation}
with homogeneous initial conditions.

Next, we introduce the energy functional
\begin{equation}
E(F)\equiv \frac{1}{2}\int \left[(\partial_{\eta} F)^2-(\partial_{r} F)^2+\frac{l(l+1)}{\sigma^2} F^2\right]dr
\end{equation}
which is conserved for $F_{0}$ but not for $f$ because we have
\begin{equation}
\frac{d}{d\eta}E(f)=\int \frac{\kappa_{5}^{4}\rho^{2}a^{2}}{18}(F_{0}+f)\partial_{\eta}f\ dr\ .
\end{equation}
Since $\sigma(r_{0}+\eta)\geq \sigma(r)$ (assuming $\sigma(r+\eta)\leq\frac{\pi}{2}$ for $k=1$) and using the Schwartz inequality together with the bounds implied by $E(f), E(F_{0})$ we arrive at the following inequality
\begin{equation}
e^{\alpha}\frac{d}{d\eta}(e^{-\alpha}\sqrt{E(f)})\leq \frac{d\alpha}{d\eta}\sqrt{E(F_{0})}\ ,
\end{equation}
where $\alpha=\int_{\eta_{0}}^{\eta}\frac{\kappa_{5}^{4}\rho^{2}a^{2}}{18\sqrt{l(l+1)}}\sigma(r_{0}+\bar{\eta})d\bar{\eta}$.

We continue by estimation of parameter $\alpha$ with the help of the behavior of functions $\sigma, H$ and the relation $\frac{\kappa_{5}^{4}\rho^2}{36} {\lesssim} H^2$ (for $k=1$ one has to assume $\frac{k}{a^2}\ll H^2$).

Finally, we obtain
\begin{equation}
\alpha\leq \sigma(r_{0}+\eta)a(\eta)\frac{H(\eta)}{2\sqrt{l(l+1)}}\equiv \beta
\end{equation}
After further manipulation we get the following estimate for the energy functional of $f$
\begin{equation}
E(f)\leq E(F_{0})(e^{\beta}-1)^2
\end{equation}

Similar estimates can be obtained for derivatives of $f$. Using Sobolev estimates the field $f$ can be controlled pointwisely provided we have initial data of compact support and $\beta\ll 1$. The last condition is satisfied only for high enough mode number $l$.

\section*{Conclusion}
We have shown that due to the modification of Friedmann equation in brane model the Huygens principle for gravitational waves seems to be satisfied only for high enough multipoles. This is in contrast with the standard cosmology. However, one should be aware that the presented analysis is something like a zero approximation of the more difficult analysis that would involve complete spectrum of brane perturbations. Here we have analyzed just the so called massless mode and reduced the brane influence to the modification of Friedmann equation.

\section*{Acknowledgements}
This work was supported by grants GACR 202/07/P284, GACR 202/09/0772 and the Czech Ministry of Education project Center of Theoretical Astrophysics LC06014.

\end{document}